\documentstyle[prl,aps,multicol]{revtex}
\input epsf

\begin{document}

\title{Sending Signals to Space-Like Separated Regions}

\author{Y. Aharonov$^{a,b}$,  and L. Vaidman$^{a,c}$}

\maketitle
\vspace{.4cm}
 
\centerline{$^a$ School of Physics and Astronomy}
\centerline{Raymond and Beverly Sackler Faculty of Exact Sciences}
\centerline{Tel-Aviv University, Tel-Aviv 69978, Israel}
\vskip .2cm
\centerline{$^b$ Physics Department, University of South Carolina}
\centerline{Columbia, South Carolina 29208, USA}
\vskip .2cm
\centerline{$^c$  Centre for Quantum Computation}
\centerline{ Department of Physics, University of Oxford,}
\centerline{ Clarendon Laboratory, Parks Road, Oxford OX1 3PU, England.}
\vskip .9cm
\centerline{Preprint requests to L.V.; E-mail: vaidman@post.tau.ac.il}

 \date{}

\vspace{.4cm}
\begin{abstract}
Two recent works suggest a possibility of sending signals to a
space-like separated region, contrary to the spirit of special
relativity. In the first work [J. Grunhaus, S. Popescu, and D. Rohrlich,
Phys. Rev. A {\bf 53}, 3781 (1996)] it has been shown that sending signals 
to particular union of space-like separated region cannot cause
causality paradoxes. Another work [Y. Aharonov and L. Vaidman,
Phys. Rev. A {\bf 61}, 052108 (2000)] showed that the relative phase of
quantum superposition of a particle in two separate locations can be
measured locally. Together with the possibility of changing the
relative phase in a nonlocal way using potential effect we,
apparently, have a method of sending signals to space-like separated 
regions. These arguments are critically analyzed in this paper.
\end{abstract}
\vspace {.5cm}

\begin{multicols}{2}

\section{Introduction}
\label{int}

Consider three space-like separated regions $A$, $B$ and $O$, Fig. 1.  Assume that
we are in $O$ and we want to send a signal to $A$ and $B$. Obviously we can send signal 
neither to $A$ nor to $B$. If we can, it will mean that in some
Lorentz frame the signal is received {\it before} it was sent. Such
causality paradox is so robust that it has no right to be considered
even in the conference on paradoxes.

\begin{center} \leavevmode \epsfbox{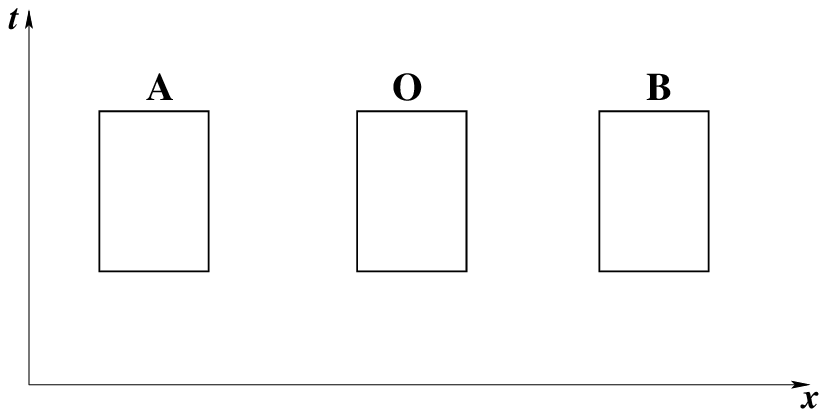} \end{center}

\noindent 
{\small {\bf Fig. 1.} A space-time diagram of space-like separated regions
  $A$, $B$, and $O$. Sending signals from $O$ to $A\cup B$ is considered.}

\vskip .4cm

 But let us ask less trivial  question: 
is it possible to send the signal from $O$ to the union of regions
$A\cup B$? Let us spell out what we mean by that. Can we make an
operation in $O$ which will lead to an  observable change in $A\cup
B$? An observer in $A$ alone as well as the observer in $B$ alone will 
not be able to observe the change, but the information they both get
does represent the change. Any local observer might know it only after 
the information from the other site will reach him, but these peaces
of information are created and irreversible recorded in $A$ and
$B$.

The simplest implementation of this situation is creation of a
random bit in $A$ and another random bit in $B$. Since the bits are
``random'', changing them does not change the information each of them
contains: random bits contain no information. However, these bits might be correlated or
anti-correlated. The parity of these bits is the information which {\it is}
stored in  the union $A\cup B$. This information might exist in the
union even when no information is contained in $A$ and $B$ separately.

Sending signals of this kind from $O$ to $A\cup B$ does not lead to
causality paradoxes: there is no Lorentz frame in which the signal is
arrived before it was sent. In every frame the signal was sent before
some  part of the information encoding the arrived signal was actualized. 

Recently, Grunhaus, Popescu and Rohrlich analyzed similar situation
\cite{GPR}. They considered ``jamming of nonlocal quantum
correlations''. This ``jamming'' is exactly this kind of producing
observable change in the union of space-like separated regions.  They derived a
simple criterion for jamming which does not lead to causality paradoxes:
the overlap of the future cones of $A$ and $B$ has to be inside the
future cone of $O$, the region of the operation of the jammer. The regions $A$, $B$ and $O$ shown in Fig. 1. obviously fulfill this criterion.

Quantum theory does allow some kind of an instantaneous change. If we
have an EPR pair of spin-$1\over2$ particles one is located in $A$ and 
another located in $B$
 \begin{equation}
\label{sing}
 |\Psi\rangle_{EPR} = {1\over {\sqrt 2}}(  |{\uparrow}\rangle_A
|{\downarrow}\rangle_B -   |{\downarrow}\rangle_A
|{\uparrow}\rangle_B),  
\end{equation}
 then the spin measurement in $B$ will change the
state of the spin in $A$ from a mixture to a pure state. However, 
the information content of $A$ is not changed. The pure state which is 
created cannot be fixed at $B$. What is fixed is the choice between
two states: one of them is created at random. Thus, a statistical
mixture is created at $A$ out of the quantum mixture by the operation at $B$. But 
the information content of this statistical mixture is equal to the
information content of the quantum mixture (it is zero, for the EPR case).

So the question is: ``Can we, in the framework of quantum theory, 
make an observable change in the union $A\cup B$?''  If we can, it
will not lead to any causality paradox. Still, it will be somewhat
paradoxical, since the spirit of special relativity tells us that
there should be no ``action at a distance'' i.e., one cannot cause a
change in a space-like separated region. The change discussed in the
preceding paragraph is not so problematic.  In the framework of the
many-worlds interpretation there is no change at all: the statistical
mixture in a particular world corresponds to the quantum mixture in
the context of the quantum state of the Universe. Performing spin
measurement in $B$ on a particle from the EPR pair will not change
anything about the particle in $A$: it will remain to be in a mixed
quantum state. The change in the correlation in the union $A\cup B$ we
discuss here is of a different type: there is no interpretation
according to which no change takes place.

Our recent work on non-locality of a quantum wave \cite{nonl} might
suggest that, nevertheless, the quantum theory, does allow
transmission of such signals. We have found that a quantum wave of a
single boson in a superposition of being in $A$ and $B$ leads to 
the EPR-type correlations between the two sites. These correlations
are governed by the relative phase between the two parts in the
superposition. But this relative phase can be changed in a non-local
way through potential effect such as the Aharonov-Bohm effect. The
effect can be generated by action in a space-like separated region
$O$. Therefore, it seems that one can cause an observable change in the
union of
space-like separated regions!

In the next section we describe the framework in which the
measurements of a quantum wave are considered. Section \ref{spnon}
describes a gedanken experiment performed on a single photon. Section \ref{spnonr}
describes a proposal for a realistic  experiment performed on a single photon and discusses the general structure of this method. Section \ref{sbnon}
describes application of the method for  a single charged
boson.  Section \ref{chan} analyzes changing of the relative
phase of the quantum state of the boson and, finally, in Section \ref{fin}
the paradox is resolved.

\section{The framework of the analysis}
\label{frame0}

We consider  a quantum wave which is an equal-weights superposition of
two localized wave packets in two separate locations:
 \begin{equation}
\label{qw}
|\Psi \rangle = 
{1\over \sqrt 2} (    | a\rangle +
~e^{i\phi} | b \rangle).
\end{equation}
 We will analyze various simultaneous (in a particular
 Lorentz frame)
measurements performed in these two locations; see Fig. 2.  
We will denote by $A$ and $B$ the space-time regions of these measurements.
The wave packet  $| a\rangle$ is localized inside the spatial region of $A$
and the wave packet  $| b\rangle$  is localized inside the spatial region of $B$.

\begin{center} \leavevmode \epsfbox{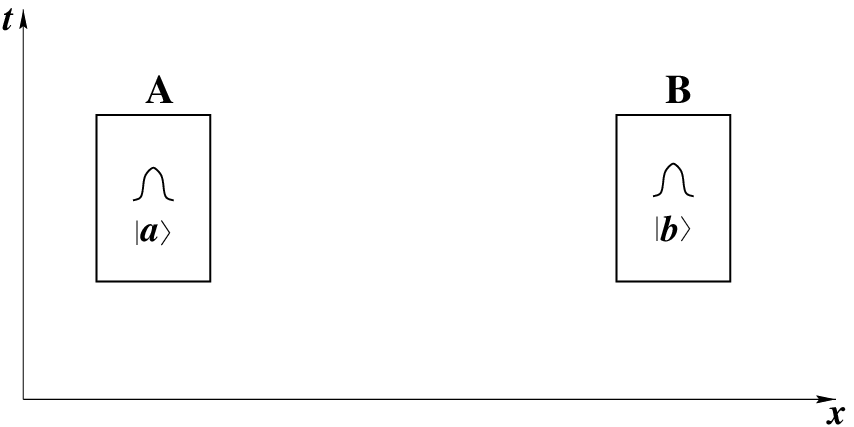} \end{center}

\noindent 
{\small {\bf Fig. 2.} Space-time diagram of the measurements performed
  on the quantum wave (\ref{qw}).}

\vskip .4cm

In order to be able to make our  analysis we have to specify exactly
the meaning of space-time regions $A$ and $B$. Are the positions of $A$ and
$B$ fixed relative to each other or are they fixed relative to an external
reference frame? Are there fixed directions in $A$ and $B$ such that
measuring devices can be aligned according to them? Is the time in
$A$ and $B$ defined relative to local clocks, or relative to an external
clock?  What are the measuring devices which are available in $A$ and $B$?
All these questions are relevant. We have to specify what is given in
$A$ and $B$ prior to bringing the quantum wave there in order to
distinguish effects related to the quantum wave from the effects
arising from our preparation and/or definition of the sites $A$ and $B$.

We make the following assumptions:

(i) There is an external inertial frame which is massive enough so
that it can be considered classical.

(ii) There is no prior entanglement of physical systems between the
sites $A$ and $B$. The two laboratories in $A$ and $B$ are also massive enough
so that the measurements performed on the quantum wave can be
considered measurements performed with classical apparatuses. However,
for various aspects of our analysis we will have to consider the two
laboratories as quantum systems. We assume that relative to the
external reference frame the two laboratories are initially described by a
product quantum state $|\Psi_A\rangle |\Psi_B\rangle$.

(iii) There is no entanglement between location of the apparatuses in $A$
and  the wave packet $|a\rangle$ (as well as between location of the
apparatuses in $B$ and the wave packet $|b\rangle$). Instead, the fact that
apparatus $A$ measures  $|a\rangle$ and apparatus $B$ measures  $|b\rangle$ is
achieved via localization relative to the external frame. The measuring
devices and the wave packets  are well
localized at the same place. This can
be expressed in the equations
\begin{eqnarray}
  \label{xx}
  \langle a|\hat x|a \rangle =  \langle \hat x_{MD_{A}} \rangle, \\
  \langle b|\hat x|b \rangle =  \langle \hat x_{MD_{B}} \rangle, 
\end{eqnarray}
where $ x_{MD_{A}}$ ($ x_{MD_{B}}$) are the variables which describe the location of
the interaction region of the measuring devices in $A$ (in $B$). It is
assumed that the wave packet $|a\rangle$ remains in the space region $A$
(and $|b\rangle$ remains in $B$) during the time of measurements.

(iv) Measurements in $A$ and in $B$ are performed by local measuring
devices activated by  {\em local} clocks, say, at
the internal time $\tau_A =\tau_B =0$. The clocks are well
synchronized with the time $t$ of the external (classical) clock:
\begin{equation}
  \label{tt}
  \langle \tau_A(t) \rangle =  \langle \tau_B(t) \rangle = t,
\end{equation}
and the spreads of the clock pointer variables $\Delta \tau_A,~\Delta
\tau_B$ are small during the experiment. Again, as stated in (ii),
there is no entanglement between clocks in $A$ and in $B$.

The assumptions can be summarized as follows: a measurement in $A$, the
space-time point relative to an external classical frame, means
a measurement performed by local apparatuses in $A$ triggered by the local
clock. The apparatuses and clocks in $A$ are not entangled with the
apparatuses and the clocks in $B$.

\section{Single-photon non-locality: a~gedanken experiment}
\label{spnon}

Let us start with considering a photon in a state (\ref{qw}).  The photon in a state (\ref{qw}) exhibits non-locality of
the EPR correlations.  The state of
the photon (\ref{qw}) can be written in the form: 
 \begin{equation}
\label{qw2}
|\Psi \rangle = 
{1\over \sqrt 2}\ (  | 1\rangle_A |0\rangle_B +
~e^{i\phi} \,| 0 \rangle_A |1\rangle_B ) ,
\end{equation}
where $ | 1\rangle_A\equiv |a\rangle$ and $ | 1\rangle_B\equiv
|b\rangle$. This form shows explicitly the isomorphism with the EPR
state (\ref{sing}).

In order to get the EPR-type correlations we must be able to perform
measurements on the photon analogous to the spin measurements in
arbitrary direction. The analog of the spin measurement in the $\hat
z$ direction is trivial: it is observing the presence of the photon in
a particular location.  A gedanken experiment yielding the analog of
the spin measurements on the EPR pair in arbitrary directions is as
follows \cite{V-sph}.  Let us consider, in addition to the photon, a
pair of spin$-{1\over 2}$ particles, one located in $A$ and one in $B$;
see Fig. 3. Both particles are originally in a spin ``down'' state in
the $\hat z$ direction. In the locations $A$ and $B$ there are magnetic
fields in the $\hat z$ direction such that the energy difference
between the ``up'' and ``down'' states equals exactly the energy of
the photon.  Then we construct a physical mechanism of absorption and
emission of the photon by the spin which is described by the unitary
transformation in each site:

 \begin{eqnarray}
 \nonumber
|1\rangle |{\downarrow}\rangle \leftrightarrow |0\rangle
 |{\uparrow}\rangle,\\
 |1\rangle |{\uparrow}\rangle \leftrightarrow |1\rangle
 |{\uparrow}\rangle,\\
\nonumber |0\rangle |{\downarrow}\rangle \leftrightarrow |0\rangle
 |{\downarrow}\rangle.
 \end{eqnarray}

\begin{center} \leavevmode \epsfbox{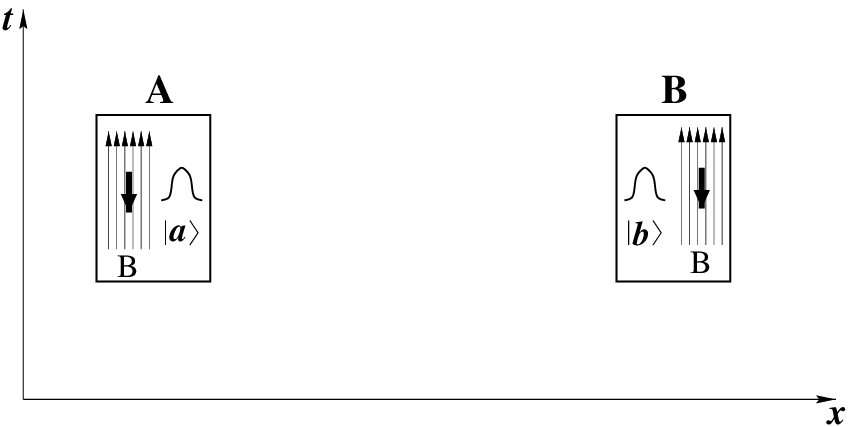} \end{center}

\noindent 
{\small {\bf Fig. 3.} Swapping of the single-photon state with the
  entangled state of two spin$-{1\over 2}$ particles.}

\vskip .8cm
\noindent
This transformation swaps the quantum state of the photon and the
quantum state of the pair of    spin$-{1\over 2}$ particles as follows:
 \begin{eqnarray}
\label{swap1} 
\nonumber
{1\over \sqrt 2}\ (  | 1\rangle_A |0\rangle_B +
e^{i\phi} \,| 0 \rangle_A |1\rangle_B ) \ |{\downarrow}\rangle_A
|{\downarrow}\rangle_B  \rightarrow
~~~~~ \\{1\over \sqrt 2}\  | 0\rangle_A |0\rangle_B \ (  |{\uparrow}\rangle_A
|{\downarrow}\rangle_B   +
e^{i\phi} |{\downarrow}\rangle_A  |{\uparrow}\rangle_B)
   .
\end{eqnarray}
Thus, we can obtain nonlocal correlations of the EPR state starting
with a single  photon, swapping its state to the state of the pair of  spin$-{1\over 2}$ particles,
 and then making appropriate spin component
measurements. Statistical analysis of the correlations between the
results of spin measurements in $A$ and in $B$ 
 allows us to find the phase $\phi$. For example, the probabilities for
 coincidence and anti-coincidence in the $x$ spin measurements are
 given by 
 \begin{eqnarray}
\label{prob-spin}
{\rm prob} (  |{\uparrow}_x \rangle  |{\uparrow}_x  \rangle) =
 {\rm prob} ( | {\downarrow}_x \rangle
|{\downarrow}_x \rangle) = {1\over 4} |1 + e^{i\phi}|^2 ,\\
{\rm prob} ( | {\uparrow}_x \rangle |{\downarrow}_x \rangle) =
 {\rm prob} ( |{\downarrow}_x \rangle
| {\uparrow}_x \rangle) ={1\over 4} |1 - e^{i\phi}|^2 .
\end{eqnarray}

We have shown that, in principal, the non-locality of a single photon
is equivalent to the non-locality of the EPR pair. Now we will turn to
the discussion of the possibilities of manifestation of this
non-locality  in  real experiments and will try to explore
the nature of this equivalence.

\section{Single-photon non-locality: a realistic experiment}
\label{spnonr}

We are not aware of experiments in which a spin in a magnetic field
absorbs a photon with high efficiency. However, there is an equivalent
operation which is performed in laboratories. There
have been several proposals \cite{Tan-sph,Fr-sph,Ha-sph,Ge-sph} how to
obtain quantum correlations based on such and similar systems. Recently there has been 
a very significant progress in  microwave cavity technology and there are
experiments in which Rydberg atoms which operate as two-level systems
 absorb and emit photons into a microwave
cavity with a very high efficiency \cite{Har}.
The excited state $|e\rangle$ and the ground state $|g\rangle$ of the
atom are isomorphic to $
|{\uparrow}\rangle$ and $ |{\downarrow}\rangle$ states of a
spin$-{1\over 2}$ particle. 
 For the atom, measuring the analog
of the $z$ spin component is trivial: 
 it is the test whether the atom is in the excited state or the ground
state.
 For measurements analogous to the spin
measurements in other directions there is an experimental solution too.
Using appropriate laser pulses the atom state can be ``rotated'' in the
two dimensional Hilbert space of ground and excited states in any
desired way. Thus, any two orthogonal states can be rotated to the
$|e\rangle$ and $|g\rangle$ states and, then, a measurement which
distinguishes between the ground and excited states distinguishes, in
fact, between the original orthogonal states.

The Hamiltonian which leads to the required interactions can be
written in the  following form:
\begin{equation}
  \label{ham}
  H = a^\dagger |g\rangle \langle e| + a  |e\rangle \langle g| ,
\end{equation}
 where $ a^\dagger$, $a$ are creation and annihilation operators of the
 photon. This Hamiltonian is responsible for the two needed operations. First,
 such coupling between the photon in the cavity in $A$ and the atom in $A$
 together with similar coupling in $B$ swaps the state (\ref{qw2}) to
 the state of two Rydberg atoms:
 \begin{eqnarray}
\label{swap2} 
\nonumber
{1\over \sqrt 2}\ (  | 1\rangle_A |0\rangle_B +
e^{i\phi} \,| 0 \rangle_A |1\rangle_B )\ |g\rangle_A
|g\rangle_B  \rightarrow
~~~~~ \\{1\over \sqrt 2}\  | 0\rangle_A |0\rangle_B \ (  | e\rangle_A |g\rangle_B +
e^{i\phi} | g \rangle_A |e\rangle_B)  
  .
\end{eqnarray}
The same Hamiltonian can also lead to an arbitrary  rotation of
the atomic state. To this end the atom has to be coupled to a cavity
with a {\em coherent state} of  photons, 
\begin{equation}
  \label{alp}
  |\alpha\rangle = e^{-{|\alpha|^2 \over 2}} 
\sum_{n=0}^{\infty} {{\alpha ^n} \over  \sqrt {n!}} \,|n\rangle .
\end{equation}
The phase of $\alpha$  specifies
the axis of rotation and the absolute value of $\alpha$ specifies the rate
of rotation. For example, the time evolution of  an atom starting at $t=0$ in the ground
state is:
 \begin{equation}
\label{psit}
|\Psi (t) \rangle = 
\cos (|\alpha| t) \, |g\rangle + {\alpha \over {i |\alpha|}} \sin
    (|\alpha| t)\, |e\rangle .
\end{equation}
This is correct when we make the approximation $a^\dagger
|\alpha\rangle \simeq \alpha ^\ast |\alpha\rangle$. This approximation
becomes  precise in
the limit of large $|\alpha|$ corresponding to a classical
electromagnetic field.  The Hamiltonian (\ref{ham}) is
actually implemented in laser-aided manipulations of Rydberg atoms
passing through microwave cavities.

\section{Charged-boson non-locality}
\label{sbnon}

Conceptually, the above scheme can be applied to any type of bosons
(instead of photons), even charged bosons.
An example of a (gedanken) Hamiltonian for this case describes a proton $ |p\rangle $ which creates
a neutron $|n\rangle$ by absorbing  a negatively charged ``meson'':
\begin{equation}
  \label{ham-me}
  H = a_m^\dagger |p\rangle \langle n| + a_m  |n\rangle \langle p| ,
\end{equation}
 where $ a_m^\dagger$, $a_m$ are creation and annihilation operators of the
 meson. This Hamiltonian swaps the state  of the meson (now
 written in the form (\ref{qw2})) and the
 state of the nucleon pair:
 \begin{eqnarray}
\label{swap3} 
\nonumber
{1\over \sqrt 2}\ (  | 1\rangle_A |0\rangle_B +
e^{i\phi} \,| 0 \rangle_A |1\rangle_B )\ |p\rangle_A
|p\rangle_B  \rightarrow
~~~~~ \\{1\over \sqrt 2}\  | 0\rangle_A |0\rangle_B \ (  |n\rangle_A |p\rangle_B +
e^{i\phi} |p\rangle_A |n\rangle_B)  
  .
\end{eqnarray}
Since there is no direct measurement of a superposition of proton and
neutron, we need again a procedure which rotates the superposition
states of a nucleon to neutron or proton state. This rotation requires
coherent states of mesons which would be, in this case, a coherent
superposition of states with different charge. Due to strong
electro-magnetic interaction the coherent state will decohere very
fast. This is essentially an environmentally induced ``charge
super-selection rule'' which prevents stable coherent superpositions
of states with different charge. It is important that there is no
{\em exact} charge super-selection rule which would prevent, in principle,
performing the experimental scheme presented above.  Indeed, Aharonov and
Susskind (AS) \cite{AS} proposed a method for  measuring the relative phase
between states with different charge, thus showing that there is no
exact charge super-selection rule. In their method one can measure the
phase even if the whole system (the observed particle and the
measuring device) is in an eigenstate of charge. This 
corresponds to initial entanglement between measuring devices in $A$ and $B$ and
thus will not be suitable for the present procedure. Here we assume
existence of superpositions of different charge states: only then it
is possible that the quantum state of measuring devices in $A$ and $B$ is
a product state.

There are some arguments that the total charge of the universe
is zero and therefore, we cannot have a product of coherent states of
charged particles in $A$ and in $B$. More sophisticated analysis has to be
performed: since the observable variables are only relative variables,
the final conclusion will be as in the AS paper \cite{AS}:
conceptually, there is no constraint on a measurement of the relative
phase of a charged boson, but decoherence will prevent construction of
any realistic experiment. See also very different arguments against
exact super-selection rule by Giulini \cite{Giu}.

\section{Is it possible to change the phase in a nonlocal way?}
\label{chan}

Now we can come back to the original question of this paper.  We have
shown that the phase $\phi$ for boson state (\ref{qw}) is locally
measurable. Given an ensemble of bosons
 with identical phase $\phi$ we
can generate a set of numbers (results of measurements) in $A$ and
another set of numbers in B, such that the two sets together yield
$\phi$.  Moreover, it seems that this phase can be changed non-locally
via the time-dependent (scalar) AB effect obtained by changing the
relative potential between the two parts of the wave during the time
they are separated.

For a charged particle this can be achieved by moving two large
oppositely charged parallel plates located between the wave packets;
see Fig. 4$a$. The two plates are placed originally one on top of the
other, i.e., there is no charge distribution and, therefore, there is
no electric field anywhere. The plates are then moved a distance $d$
apart and then, after short time $t$, they are brought back. We will
call such an operation ``opening a condenser''. Opening the condenser
with charge density $\sigma$ will lead to the change of the relative
phase of the particle with charge $q$ equal
\begin{equation}
  \label{phase}
  \Delta \phi = {{4\pi \sigma d q  t}\over \hbar}.  
\end{equation}
For $\Delta \phi =\pi$ the operation seemingly sends one bit of
information from $O$ to $A\cup B$, see Fig.~5.

Consider now a neutral boson state. A massive plate in between the
regions $A$ and $B$ which we move or not move towards one of the sites
will introduce the phase shift in complete analogy with the scalar AB
effect, Fig.~4$b$. The difference here is that the gravitational
fields in the regions $A$ and $B$ are not zero, but the fields are not
affected by the motion of the plate.  Thus, again, it seems that by an
action in a localized region we can send information to a space-like
separated region.  Moving or not moving the massive plate changes the
relative phase of the components of the quantum wave, and, therefore,
apparently changes correlations in the results of measurements in $A$
and $B$; see Fig. 5.

\begin{center} \leavevmode \epsfbox{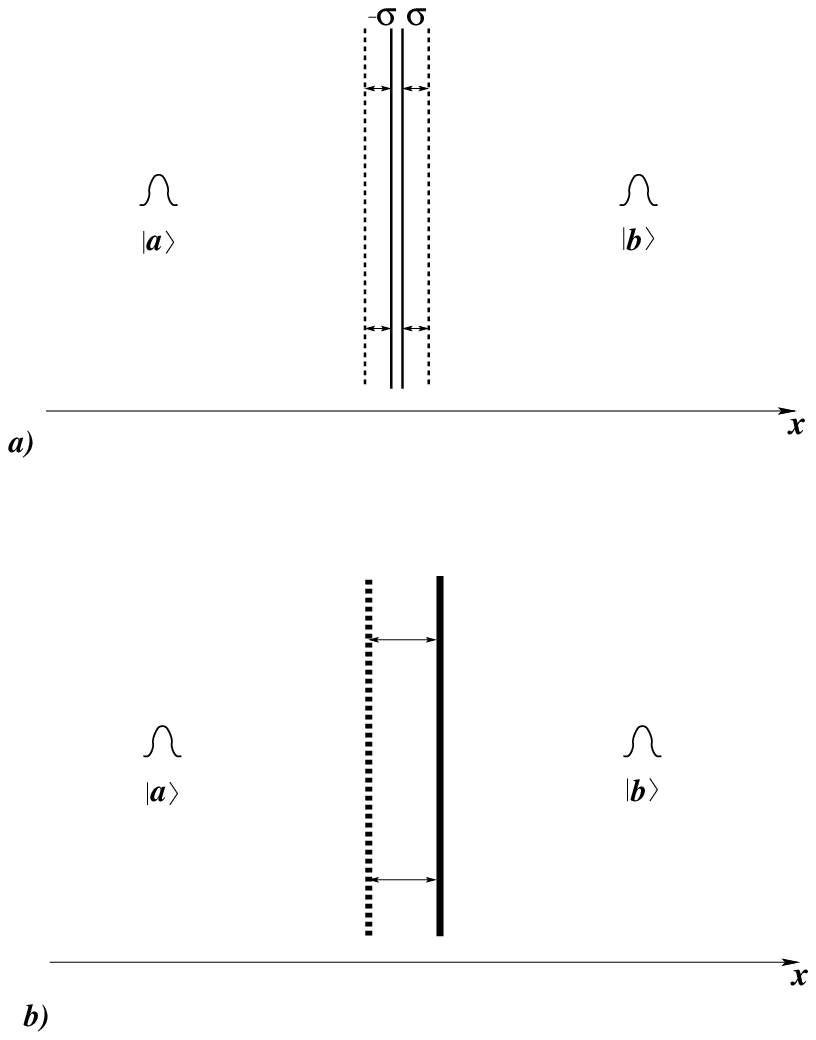} \end{center}

\noindent 
{\small {\bf Fig. 4.}  {\em a}). Parallel-plate condenser with charged
  plates, originally one on top of the other, is opened (by moving the
  plates apart) for a short time while the wave packets $| a\rangle$
  and $| b\rangle$ are far apart. This operation introduces change in
  the electric potentials between the locations of $| a\rangle$ and $|
  b\rangle$ which generates the AB phase.}  {\em b}). A massive plate
produces different gravitational potential at the locations of the
wave-packets if it is closer to one of them. The relative phase is
changed if the plate is moved  or not moved.

\vskip .4cm

\begin{center} \leavevmode \epsfbox{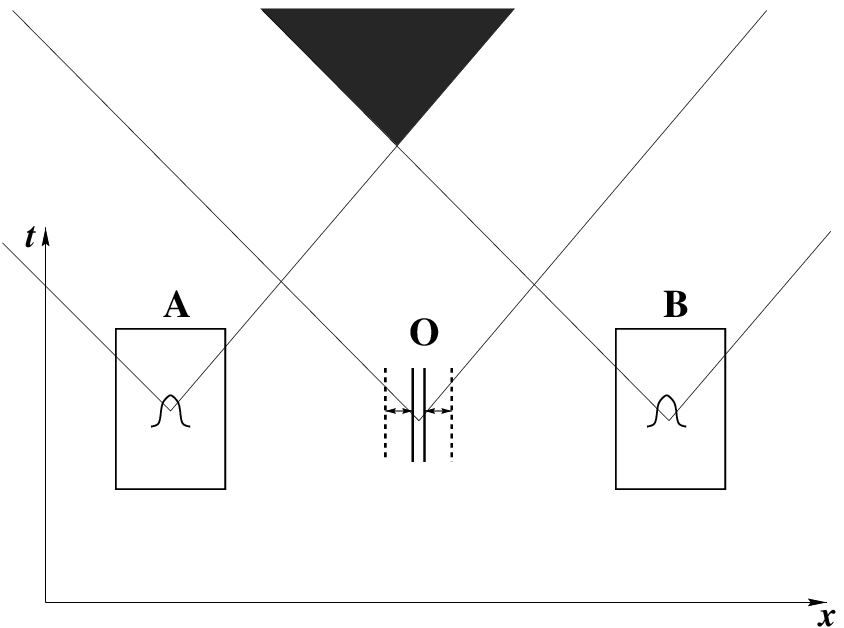} \end{center}

\noindent 
{\small {\bf Fig. 5.} Apparent sending signals to a union of space-like
  separated regions. Operation in $O$, like an opening the condenser for a period
  of time, apparently changes the correlations between measurements in
  $A$ and $B$. No signal is sent from O, neither to $A$ nor to B, but the
  signal {\em is} sent to the union of $A$ and $B$. The intersection of
  light cones originated at $A$ and at $B$ lies inside the light cone
  originated at $O$.  Therefore, the action of the condenser falls into
  the category of ``jammers'' considered in Ref.  \cite{GPR}.}

\vskip .4cm

  \section{Resolution of the paradox}
  \label{fin}

  In spite of the fact that we cannot reach a causality paradox using
  the procedure described above, it  clearly contradicts the spirit, if
  not the letter, of special relativity. And, in fact, it is
  impossible. We can cause an observable change neither in a localized 
  space-like separated region nor in the union of such regions.

 It is incorrect that the opening of a condenser will change
  correlations between results of measurements in $A$ and $B$. It must be
  incorrect because we should be able to use a covariant gauge in
  which changes in the potentials take place only inside the light
  cone. However, we can explain this phenomena also in a standard
  (Coulomb) gauge. In our
  scheme the measurements in local sites include interactions with
  coherent states of auxiliary particles,  particles which are
  identical to the particle in a superposition. Therefore, if the
  particle in question is charged, the auxiliary particles are also charged
  and opening the condenser  changes the phase of the coherent
  state in such a way that the correlations are  not  changed. The
  gauge which we choose changes the description of auxiliary particles
  too, so that the probabilities for results of measurements remain
  gauge invariant.
  
  Consider now a neutral boson state.  The resolution of the paradox
  in this case is similar to the resolution of Einstein's paradox of
  an exact energy of an exact clock \cite{ETP}.  The explanation is
  that the pointers of the local clocks are shifted.  Simultaneity
  between $A$ and $B$ is altered due to the action of the massive plate.
  Since in our case local clocks activate the measurements, the shift
  in the pointer will lead to a change. This change compensates
  exactly the phase change of the boson.

\vskip .4cm \noindent {\bf Acknowledgments}
 
  This research was supported in part by grant
 471/98 of the Basic Research Foundation (administered by the Israel
 Academy of Sciences and Humanities) the NSF grant PHY-9971005 and the 
 EPSRC grant  GR/N33058.

\end{multicols}

\end{document}